\documentstyle[preprint,aps]{revtex}
\input epsf

\begin{document}

\draft

\title{Cancellation of ladder graphs in an effective expansion}

\author{M.~E.~Carrington $^a$, R.~Kobes $^a$ and E.~Petitgirard $^{a,b}$}

\address{
   $^a$ Department of Physics, University of Winnipeg,\\
   Winnipeg, Manitoba, R3B 2E9, Canada\\
   $^b $Department of Physics, Brookhaven National Laboratory,\\
   Upton, New York, 11973-5000, USA}

\date{\today}

\maketitle

\begin{abstract}
  A resummation of ladder graphs is important in
  cases where infrared, collinear, or light--cone singularities
  render the loop expansion invalid, especially at high temperature
  where these effects are often enhanced. It has been noted in some
  recent examples of this resummation that the ladder graphs are
  canceled by other types of terms. In this note we show that
  this cancellation is quite general, and for the most part algebraic.
\end{abstract}
\vspace{3cm}
\pacs{PACS numbers: 11.10Wx, 11.15Tk, 11.55Fv}

\narrowtext

\section{Introduction}
\label{sec1}
The loop expansion for gauge theories at high temperature suffers
from a number of problems due to the extreme nature of the infrared
divergences present. To address these difficulties the hard thermal
loop expansion was devised. This expansion is an effective reordering of the
perturbation theory to take into account equivalent orders of
loop diagrams to any given order \cite{r8,klim,wel,brat,wong,tay}.
 Although successful in
resolving many paradoxes, there still remain some fundamental
problems with this expansion in certain limits outside of its
range of validity. One particular problem is that of the damping
rate of a fast fermion, where a self--consistent
calculational scheme outside of the hard thermal loop expansion
has been used \cite{smilga,piz}. Another class of such
problems involves
processes sensitive to the behaviour near the light--cone, where
an ``improved'' hard thermal loop expansion has been proposed
\cite{flech,kraemmer}.
\par
In this note we consider perturbation expansions beyond the
loop expansion which include
ladder graphs. These graphs, which are not included in the
hard thermal loop expansion, become important in certain cases
sensitive to the infrared and/or light--cone limits
\cite{smilga,kraemmer,pressure,meg}. They also arise in the
context of the eikonal expansion of gauge theories \cite{cornwall,hou}.
It has been noticed previously that in certain cases inclusion of
these ladder graphs leads to an intricate cancellation between certain
terms involving effective propagators and vertices. The purpose
of this note is to show in a relatively simple and general way how and under
what circumstances this cancellation occurs.
\section{Ladder Graphs}
\label{sec2}
\begin{figure}
\begin{center}
\leavevmode
\epsfxsize=4 in
\epsfbox{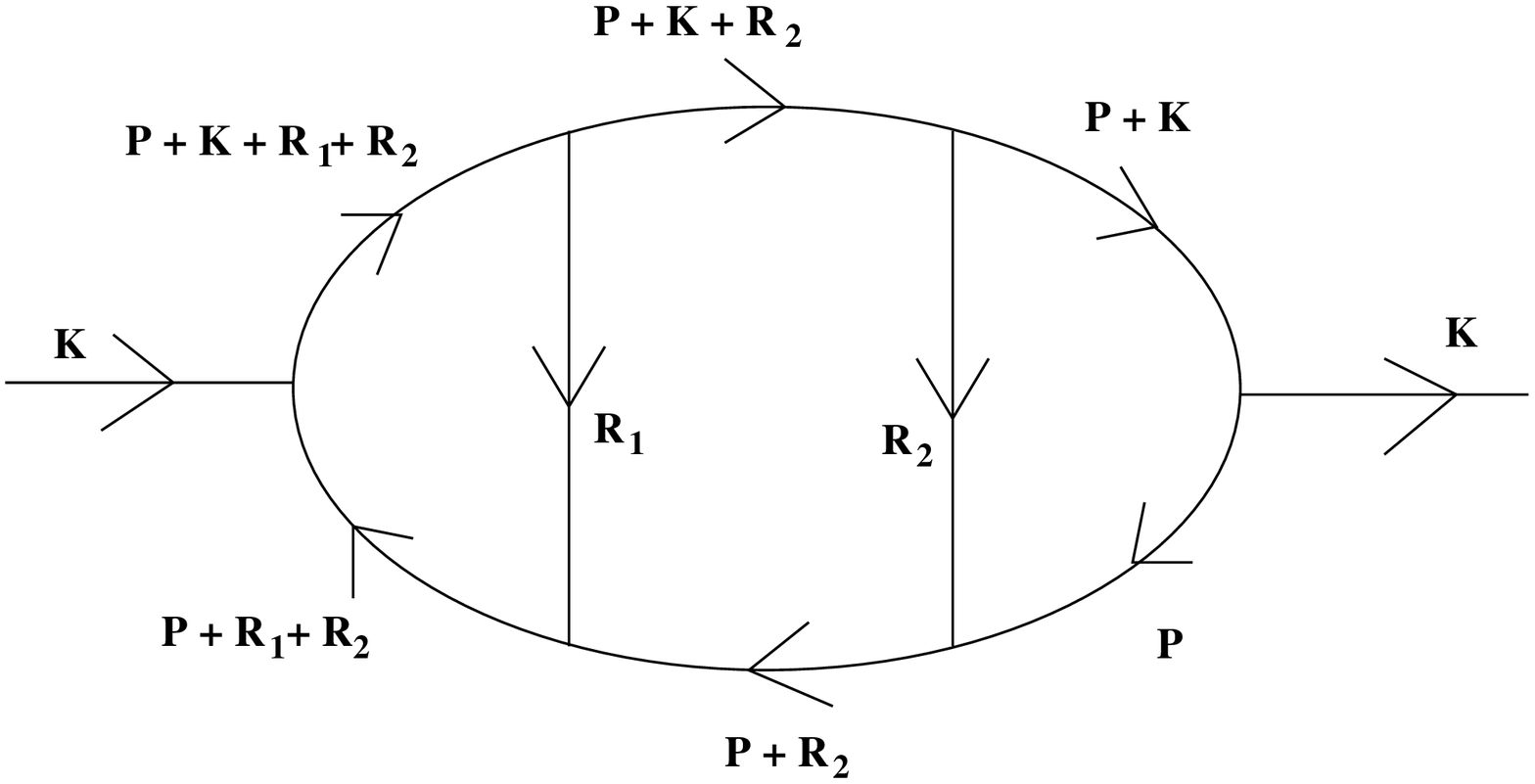}
\end{center}
\caption{A three--loop self--energy ladder graph}
\label{ladderfig}\end{figure}

In this section we show under what circumstances ladder graphs are
important, and give a method for their inclusion.
We work here with a scalar $\phi^3$ theory, but the results
generalize straightforwardly for other theories.
Consider first the graph of Fig.\ref{ladderfig}, which is given by
\begin{eqnarray}
  \label{ladder}
  -i\Sigma(K)& =& (-ig)^6\int\, dR_1\, dR_2\, dP \,
  \Delta(P+R_1+R_2)\Delta(P+R_1+R_2+K)\Delta(R_1)
  \Delta(P+R_2)\nonumber\\& &\Delta(P+R_2+K)\Delta(R_2)
  \Delta(P)\Delta(P+K),
\end{eqnarray}
where $\Delta(K) = i / (K^2+i\epsilon)$
and $K=(k_0, {\vec k})$. Such contributions are known to be important in $QED$
for instance when taking $P$ hard and $R_i$ soft \cite{smilga}. Finite
temperature effects can be handled by using the imaginary--time formalism
\cite{land}, although for our purposes a real time formalism such as the
Keldysh basis or the $R/A$ formalism is more convenient \cite{aurenche}. In
cases where the loop expansion is valid this graph would be suppressed by a
factor of $g^4$ relative to the one--loop graph of Fig.~\ref{oneloop}. However,
especially at finite temperature, circumstances could arise where this is not
the case. Let us split two of the propagators in Eq.~[\ref{ladder}] as
\begin{equation}
  \Delta(P+R_2)\Delta(P+R_2+K) = i
  \frac{[\Delta(P+R_2)-\Delta(P+R_2+K)]}{K^2+2K\cdot (P+R_2)},
\end{equation}
and furthermore consider the infrared limit $2K\cdot R_2 \ll (K^2 + 2K\cdot
P)$, whereby
this splitting is approximated by
\begin{equation}
  \label{split}
  \Delta(P+R_2)\Delta(P+R_2+K) \approx i
  \frac{[\Delta(P+R_2)-\Delta(P+R_2+K)]}{K^2+2K\cdot P}.
\end{equation}
We perform an analogous split for the product
$\Delta(P+R_1+R_2)\Delta(P+R_1+R_2+K)$.
Such approximations in Eq.~[\ref{ladder}] lead to
\begin{eqnarray}
  \label{splitladder}
  -i\Sigma(K) &\approx& (-ig)^6\int\, dR_1\, dR_2\, dP\,
  i\frac{[\Delta(P+R_1+R_2)-\Delta(P+R_1+R_2+K)]}{K^2+2K\cdot P}
  \Delta(R_1)\nonumber\\
  & &i\frac{[\Delta(P+R_2)-\Delta(P+R_2+K)]}{K^2+2K\cdot P}
  \Delta(R_2)\Delta(P)\Delta(P+K).
\end{eqnarray}
Now, if it happens that
a region of phase space exists where $(K^2+2K\cdot P)$
is sufficiently small (for example, at finite temperature,
$(K^2+2K\cdot P) \sim O(g^2T^2)$), then
a factor of $g^4$ arises in the denominator of Eq.~[\ref{splitladder}] which
would cancel a factor of $g^4$ in the numerator. This would lead  to a
situation where the ladder graph of Fig.~\ref{ladderfig} is of the
same order as the one--loop term of Fig.~\ref{oneloop}, signaling the
breakdown of the loop expansion.
\par
The same situation occurs in the light-cone limit $K^2=0$ where the region of
the phase space $P^2 \sim O(g^2T^2)$ and
$1\pm\hat{p}\cdot\hat{k} \sim O(g)$ becomes important. The previous
approximation $2K\cdot R_2 \ll (K^2 + 2K\cdot P)$ is apparently no longer
possible, but the various denominators $2K\cdot P+2K\cdot R_i$ become small
enough ($\sim O(g^2T^2)$ and below) to compensate the extra factors of $g$ in
the numerator.
\par
\begin{figure}
\begin{center}
\leavevmode
\epsfxsize=4 in
\epsfbox{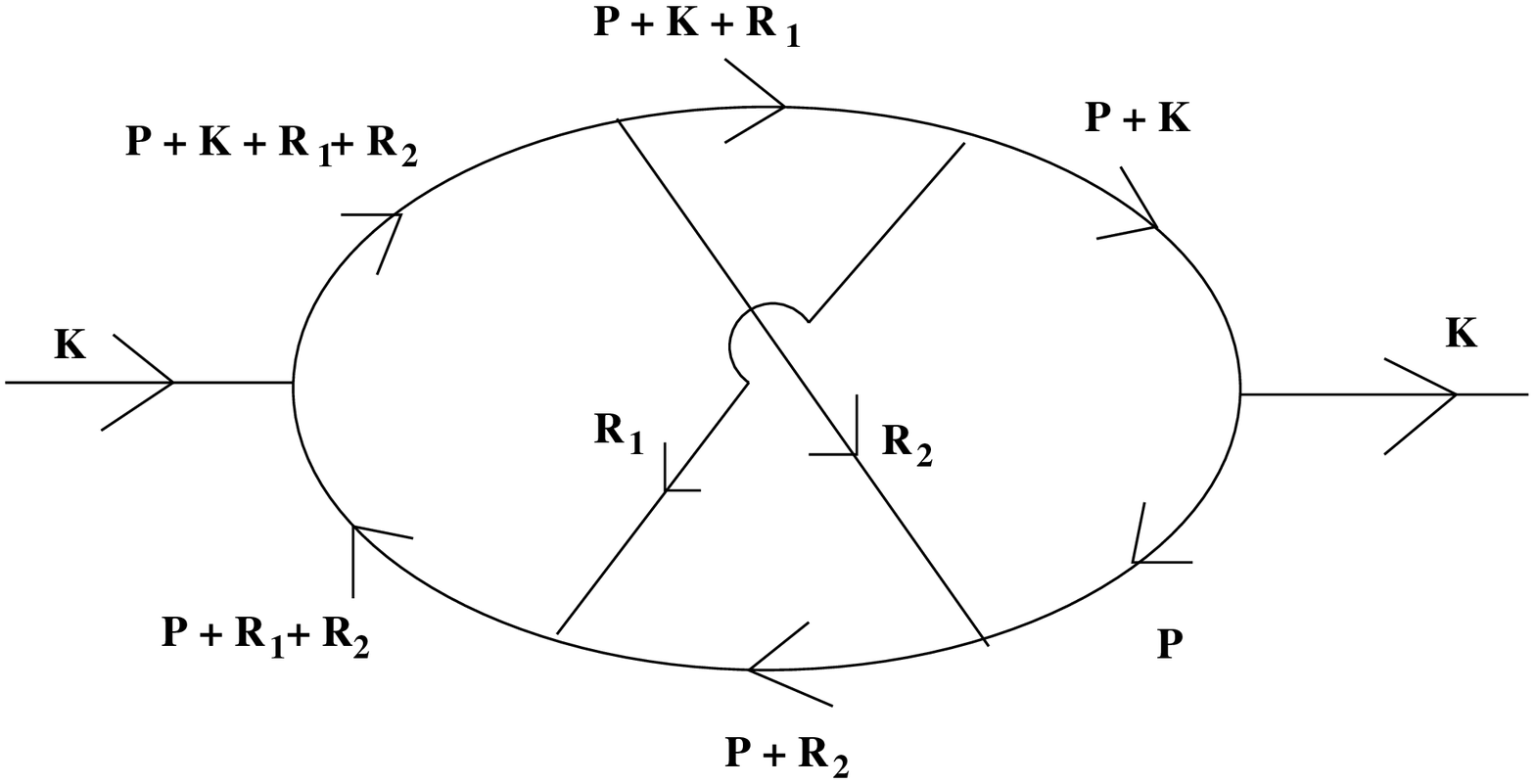}
\end{center}
\caption{A three--loop self--energy non--ladder graph}
\label{unladderfig}\end{figure}

The breakdown of the loop expansion in this manner is due to the importance
of the ladder graphs like Fig.~\ref{ladderfig} and similar higher loop terms.
Higher loop ``crossed terms'' such as that illustrated in
Fig.~\ref{unladderfig},
\begin{eqnarray}
  -i\Sigma(K)& =& (-ig)^6\int\, dR_1\, dR_2\, dP\,
  \Delta(P+R_1+R_2)\Delta(P+R_1+R_2+K)\Delta(R_1)
  \Delta(P+R_2)\nonumber\\ & &\Delta(P+R_1+K)\Delta(R_2)
  \Delta(P)\Delta(P+K),
\end{eqnarray}
do not contribute in the same way as the ladder graphs. This is because that
while
the product of propagators $\Delta(P+R_1+R_2)\Delta(P+R_1+R_2+K)$ can be split
along the lines of Eq.~[\ref{split}], the product
$\Delta(P+R_2)\Delta(P+R_1+K)$ would be split as
\begin{equation}
  \Delta(P+R_2)\Delta(P+R_1+K) =
  i\frac{[\Delta(P+R_2)-\Delta(P+R_1+K)]}
  {K^2+2K\cdot (P+R_1) + (P+R_1)^2 - (P+R_2)^2}.
\end{equation}
Due to the presence of the
$(P+R_1)^2 - (P+R_2)^2$ term,
the infrared limit $2K\cdot R_1 \ll (K^2 + 2K\cdot P)\sim O(g^2T^2)$ or the
light-cone limit $(2K\cdot P+2K\cdot R_i) \sim O(g^2T^2)$ and below would
not by themselves lead to a
cancellation of a factor of $g^2$ in the numerator . One could try to get such
a cancellation by furthermore restricting the phase space so that $P\cdot R_i$
and $R_i^2$ ($i=1,2$) is sufficiently small, but this introduces extra factors
of $g$ in the numerator coming from the momentum integral over $P$.
The conclusion one draws is that in the infrared and light-cone limits such
crossed graphs are suppressed relative to the ladder graphs.
\section{Ladder resummation}
\label{sec3}
\begin{figure}
\begin{center}
\leavevmode
\epsfxsize=4 in
\epsfbox{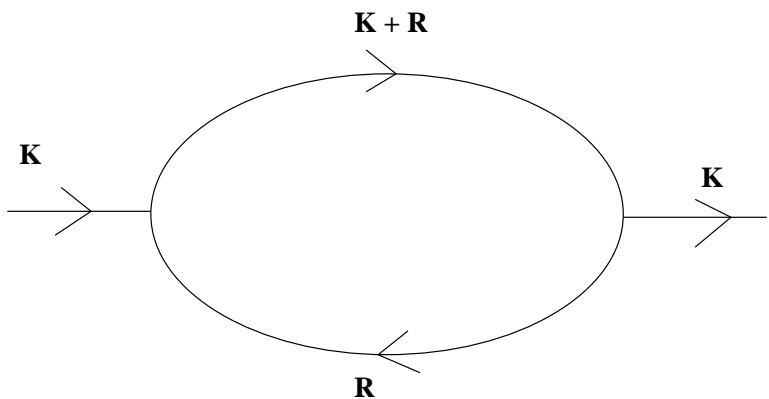}
\end{center}
\caption{A one--loop self--energy graph}
\label{oneloop}\end{figure}

\begin{figure}
\begin{center}
\leavevmode
\epsfxsize=4 in
\epsfbox{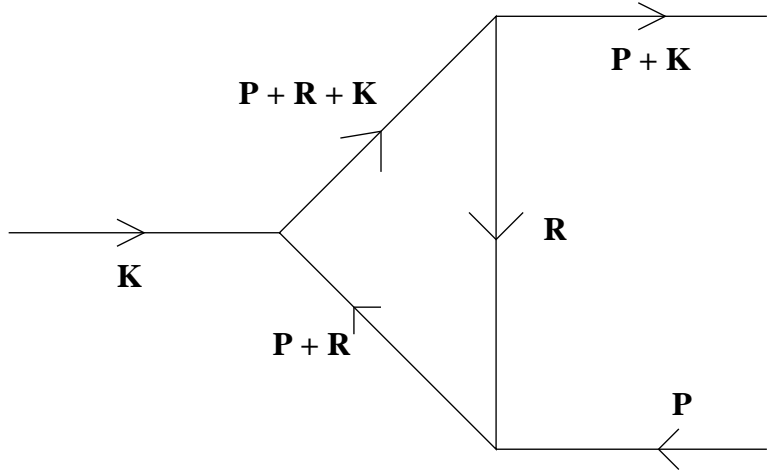}
\end{center}
\caption{A one--loop vertex graph}
\label{vertexfig}\end{figure}

In this section we describe a method for including the ladder graphs
discussed in the previous section in an effective expansion. To this
end, we first consider the one--loop vertex of Fig.~\ref{vertexfig}.
The expression for this graph is
\begin{equation}
  -i\Gamma(K, P) = (-ig)^3 \int\, dR\, \Delta(R)\Delta(R+P)\Delta(K+P+R).
\end{equation}
We split the two propagators $\Delta(R+P)\Delta(K+P+R)$ as in Eq.~[\ref{split}]
and
use the approximation $2K\cdot R \ll (K^2 + 2K\cdot P)$, whereby this equation
becomes
\begin{equation}
  -i\Gamma(K, P) \approx (-ig)^3 \frac{i}{K^2+2K\cdot P}
  \int dR \Delta(R)\left[\Delta(R+P)-\Delta(K+P+R)\right].
\end{equation}
Comparing this to the one--loop self--energy graph of Fig.~\ref{oneloop}:
\begin{equation}
  \label{oneloopselfenergy}
  -i\Sigma(P) = (-ig)^2 \int\, dR\, \Delta(R)\Delta(R+P),
\end{equation}
we find the relation
\begin{equation}
  \label{relation}
  \Gamma(K, P) \approx g\frac{1}{K^2+2K\cdot P}
  \left[\Sigma(P) -\Sigma(K+P)\right].
\end{equation}
Note that, due to the absence of an $i\epsilon$ in the
denominator, we must assume
that we  are in a region of phase space where $K^2+2K\cdot P$
does not vanish.
\par
\begin{figure}
\begin{center}
\leavevmode
\epsfxsize=5 in
\epsfbox{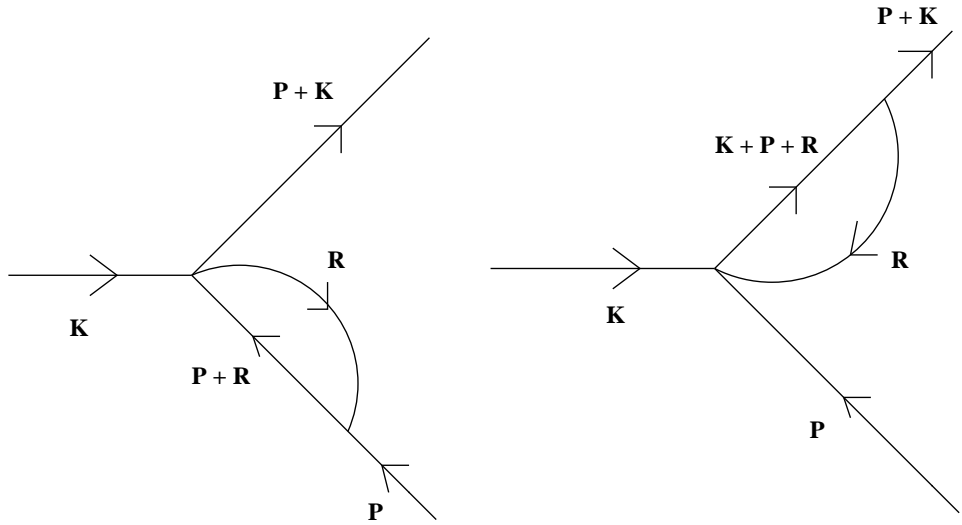}
\end{center}
\caption{Extra contributions in addition to
 Fig.~\ref{vertexfig} to the one--loop 3--point
scalar--photon vertex in scalar $QED$}
\label{sqedvertex}\end{figure}

Results similar to Eq.~[\ref{relation}] hold in gauge theories. For
example, in scalar $QED$, to one--loop the three--point scalar--photon
vertex is given by the graph of Fig.~\ref{vertexfig} together with the
extra contributions of Fig.~\ref{sqedvertex} (in these graphs, the
photons are the lines with momentum $K$ and $R$).
Splitting the propagators as in Eq.~[\ref{split}] and imposing the
limit $2K\cdot R \ll (K^2+2K\cdot P)$, we find the vertex can be written as
\cite{meg}
\begin{equation}
  \label{sqedrelation}
  \Gamma_\mu(K, P) \approx g\frac{K_\mu + 2P_\mu}{K^2+2K\cdot P}
  \left[\Sigma(P) -\Sigma(K+P)\right],
\end{equation}
where the one--loop self--energy graph is shown in Fig.~\ref{oneloop}
(with the photon line having momentum $R$).
This relation illustrates the connection between this approximation for
the vertex function and gauge invariance: contracting
Eq.~[\ref{sqedrelation}] with $K^\mu$ leads to
\begin{equation}
  K\cdot \Gamma(K, P) = g  \left[\Sigma(P) -\Sigma(K+P)\right],
\end{equation}
which of course is the Ward identity for the vertex function of scalar
$QED$. In a sense, this approximation for the vertex function is
equivalent to ``solving'' the Ward identity for this function.
Similar relations and conclusions can be made for ordinary $QED$ with fermions.
\par
\begin{figure}
\begin{center}
\leavevmode
\epsfxsize=4.5 in
\epsfbox{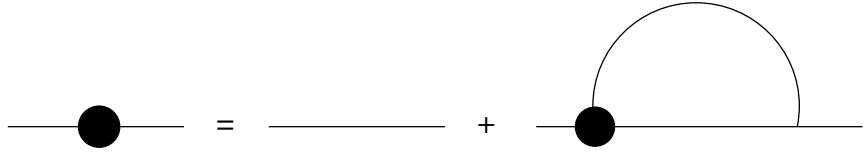}
\end{center}
\caption{A partial Schwinger--Dyson equation for the full self--energy
  which generates the ladder graph resummation}
\label{semiselfenergy}\end{figure}
\begin{figure}
\begin{center}
\leavevmode
\epsfxsize=4.5 in
\epsfbox{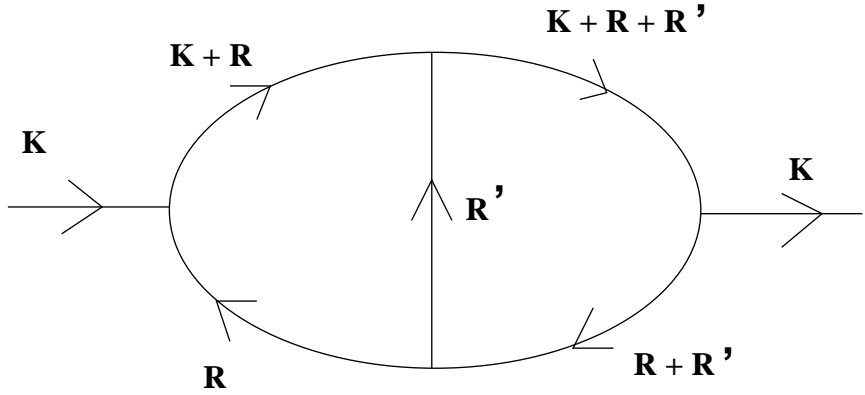}
\end{center}
\caption{A two--loop ladder graph contribution to the self--energy}
\label{firstladder}\end{figure}

Although Eq.~[\ref{relation}] was derived for one--loop values,
the same general form holds at higher orders under the equivalent
approximations. Alternatively, one could view the relation
of Eq.~[\ref{relation}] as
an approximate solution to the Schwinger--Dyson equation for
the full vertex function. It is in this sense that this relation
for the vertex function can be seen to generate a resummation of
ladder graphs. Consider the partial Schwinger--Dyson equation for the
full self--energy indicated in Fig.~\ref{semiselfenergy},
\begin{equation}
  \label{semisdequation}
  -i\Sigma(K) = -ig\int\, dR\,\left[-i\Gamma(K, R)\right]
  \Delta(R)\Delta(K+R),
\end{equation}
where $\Gamma(K,R)$ is the full three--point vertex.
Iterating this equation using Eq.~(\ref{relation}) with an appropriate
$\Sigma$ is seen to generate the perturbative summation of ladder
graphs, such as in Fig.~\ref{ladderfig}, under the appropriate
approximations of small loop momenta used in the derivation.
For example, if for the first iteration we use $\Gamma(K,R)$ of
Eq.~(\ref{relation}) with the one--loop self--energy of
Fig.~\ref{oneloop} given in Eq.~(\ref{oneloopselfenergy}),
we find
\begin{eqnarray}
   -i\Sigma(K)& =& -(ig)^2\int\, dR\,
   \frac{\left[\Sigma(R) -\Sigma(K+R)\right]}{K^2+2K\cdot P}
   \Delta(R)\Delta(K+R)\nonumber\\
   &=& -(ig)^2\int\, dR\,\frac{ \Delta(R)\Delta(K+R)}{K^2+2K\cdot R}
   i(-ig)^2\int\,dR^\prime\,\nonumber\\
   & &\qquad\left[
     \Delta(R^\prime)\Delta(R+R^\prime)-
     \Delta(R^\prime)\Delta(K+R+R^\prime)\right]\nonumber\\
   &=&(-ig)^4\int\,dR\,dR^\prime\,
   \Delta(R)\Delta(K+R)\Delta(R^\prime)\Delta(R+R^\prime)
   \Delta(K+R+R^\prime),
\end{eqnarray}
where we assumed $2K\cdot R^\prime \ll (K^2 + 2K\cdot R)$.
This expression corresponds to the ladder graph of
Fig.~\ref{firstladder}.
\par
In the light-cone limit $K^2=0$ and $P^2\sim O(g^2T^2)$, one 
can show that a form similar to Eq.~(\ref{relation})
can be obtained, provided that the external momenta $P\cdot K$ is restricted
to lie between $O(gT^2)$ and $O(g^2T^2)$ but not below.
In that case, it turns out that
the main contribution comes from the delta functions arising in
$\Delta(P+R)$ and $\Delta(P+R+K)$ or their Breit-Wigner
counterparts if the full propagators
are used (see the following section). The restrictions imposed allow us to
discard the terms $2K\cdot R$ as
negligible compared to $2P\cdot K$ and to get a similar form as
 that of Eq.~(\ref{relation}). In particular, in $QED$, one has
 \begin{equation}
   \Gamma_\mu(K, P) \approx g\frac{P_\mu}{K\cdot P}
  \left[\Sigma(P) -\Sigma(K+P)\right],
\end{equation}
where it is understood that the phase space is restricted to the
appropriate region.
\section{Mechanism of cancellation of ladder terms}
\label{sec4}
Previous works where a resummation of ladder graphs has been used
have noticed that there is a cancellation of the contributions of the ladder
graphs when effective propagators are used on the internal lines
of Fig.~\ref{semiselfenergy}. This effect has
been seen in the calculation of the fast fermion damping rate when
damping effects of the photon/gluon have been included \cite{smilga}, and
also in the calculation of the self--energy in scalar $QED$
\cite{kraemmer,meg}, again with damping effects included.
This cancellation was shown by choosing
a particular form of $\Sigma(K)$ to include damping effects, and
then iterating a Schwinger--Dyson type of equation. In these works
it was not apparent if this cancellation was the result of the
particular choice of $\Sigma(K)$ used. In this section we show that
such a cancellation occurs quite generally and, to a large
extent, algebraically.
\par
\begin{figure}
\begin{center}
\leavevmode
\epsfxsize=4.5 in
\epsfbox{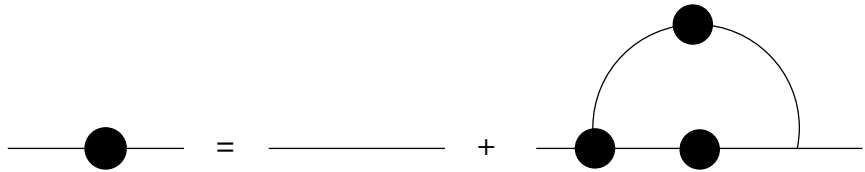}
\end{center}
\caption{The Schwinger--Dyson equation for the full self--energy}
\label{fullselfenergy}\end{figure}

We begin by noting that the ladder resummation generated by the
partial Schwinger--Dyson equation of Eq.~(\ref{semisdequation}) is
incomplete; one must include the effects of the self--energy
corrections on the internal lines, as in Fig.~\ref{fullselfenergy}.
This was emphasized in Refs.~\cite{smilga,kraemmer,meg} from the
point of view of gauge invariance. We thus consider the full
Schwinger--Dyson equation of Fig.~\ref{fullselfenergy}:
\begin{equation}
  \label{sdequation}
  -i\Sigma(K) = -ig\int\, dR\,\left[-i\Gamma(K, R)\right]
  G(R)G(K+R),
\end{equation}
where $\Gamma(K,R)$ is the full three--point vertex and
$G(K) = i / (K^2-\Sigma(K)+i\epsilon)$
is the full propagator.
We insert for the full vertex function on the right--hand--side of
this equation the tree--level value plus
the effective vertex of Eq.~[\ref{relation}]:
\begin{equation}
  -i\Sigma(K) =(-ig)^2\int \,dR\,
  \left[1+\frac{\left[\Sigma(R)-\Sigma(K+R)\right]}
    {K^2+2K\cdot R}\right]G(R)G(K+R).
\end{equation}
This can subsequently be rewritten as
\begin{equation}
  -i\Sigma(K) =(-ig)^2\int\, dR\,
  \frac{i}{K^2+2K\cdot R} \left[G(R)-G(K+R)\right].
\end{equation}
Shifting variables in the second term, and using the relation
$G(-R)=G(R)$, we find
\begin{equation}
  \label{resummed}
  -i\Sigma(K)= 2(-ig)^2\int\, dR\, \left[\frac{i}{K^2+2K\cdot R}\right]
  \left[\frac{i}{R^2-\Sigma(R)+i\epsilon}\right].
\end{equation}
We recall in this derivation that we have assumed
$K^2+2K\cdot R$ does not vanish, so that the integral is
well defined.
\par
We compare the result of Eq.[\ref{resummed}] to that obtained
without the use of effective propagators and vertices:
\begin{equation}
  \label{bare}
  -i\Sigma(K) = (-ig)^2\int\, dR\,
  \left[\frac{i}{(K+R)^2 +i\epsilon}\right]
  \left[\frac{i}{R^2+i\epsilon}\right].
\end{equation}
An explicit comparison depends on the form of the self--energy
function $\Sigma(R)$ in Eq.~[\ref{resummed}], but in the general
case that the contribution of $\Sigma(R)$ to the pole of
$1/[R^2-\Sigma(R)]$ can be neglected, one can show that
the two results of Eqs.[\ref{resummed}, \ref{bare}] are equivalent.
One can see this loosely as follows. Eq.~(\ref{bare})
has two contributions: one from $R^2=0$ and one from
$R^2+K^2+2K\cdot R = 0$. Performing the contour integration,
and recalling that we have assumed $K^2+2K\cdot R$ does not
vanish (which is necessary for the ladder graphs to contribute in
the first place), we find the result to be equivalent to
Eq.~(\ref{resummed}), assuming the effects of $\Sigma(R)$ can
be neglected in the integration of Eq.~(\ref{resummed}).
\par
In particular, in the light-cone limit in QED, and still under the
restriction on the phase-space that $R\cdot K$ lies between
$O(gT^2)$ and $O(g^2T^2)$, one finds for the (retarded) polarization
tensor at leading order ($R=K+P$)
 \begin{eqnarray}
    i\Pi^{00}_R(K) &=&-(-ig)^2\int \, dR\, 4p^2\frac{i}{K\cdot R}
   \left[ \left({1\over 2}-n_F(r_0)\right) \left[\frac{i}{R^2-
         {1\over2}{\rm Tr}(\gamma\cdot R\,\,\Sigma_R(R))}\right] \right.
    \nonumber\\
   &+& \left. \left({1\over 2}-n_F(p_0)\right) \left[
       \frac{i}{P^2-{1\over 2}{\rm Tr}
         (\gamma\cdot P\,\,\Sigma_A(P))}\right] \right]
\end{eqnarray}
where taking a constant damping rate for $\Sigma$, for instance,
doesn't give rise to any contribution to the integral
and leads to the usual (and partial) hard thermal loop result.
We note that this restriction on $\Sigma$
does not allow us to investigate the scale where the
effect of an asymptotic mass for the fermion might
become relevant\cite{flech,kraemmer}, 
which is beyond the scope of this paper.
\par
One can interpret this kind of cancellation as occurring between
self--energy and vertex corrections in the original Schwinger--Dyson
equation of Eq.~[\ref{sdequation}], as was found explicitly in
Refs.\cite{smilga,kraemmer,meg}. We see by these
methods that such a cancellation is to a large extent algebraic,
depending ultimately on the question of the contribution of
the self--energy function $\Sigma(R)$ to the poles of the
propagator. This method of cancellation
thus has a weaker dependence on the particular form of $\Sigma$ used
than was employed in Refs.\cite{smilga,kraemmer,meg};
in particular, it is possible that self--energy functions with
a non--trivial momentum dependence could be used with the same
method of cancellation at work.
\section{Conclusions}
\label{sec5}
We have considered an effective expansion based on an approximate
solution of the Schwinger--Dyson equation for the full self--energy
function which generates, among other terms, a resummation of
ladder graphs. Such graphs, especially at finite temperature, are
important when the loop expansion starts to fail due to extreme
behaviour in the infrared, collinear, or light--cone limits. However, as was
shown in some specific examples \cite{smilga,kraemmer,meg}, the
effects of the ladder graphs cancel against certain self--energy
insertions. We have given a relatively simple proof of
this fact which is for the most part algebraic, the only additional
input being an assumption on the relative contribution
of the self--energy insertion to a contour integral.
\par
Although fairly general, this proof
does not preclude the possibility that ladder graphs may
contribute in other contexts. In particular, one might encounter
cases where the basic assumption on the ``small''
internal loop momenta $R$ does not satisfy
$2K\cdot R \ll (K^2+2K\cdot P)$ for some external momenta $K$ and $P$.
Such cases may arise, again at finite temperature, in some combination
of an extreme infrared, light--cone and/or collinear limit, and would
signal the breakdown of the basic relation of Eq.~(\ref{relation})
used here to ``solve'' the Schwinger--Dyson equation for the vertex
in terms of the self--energy function. It is not known if a cancellation
between ladder graphs and self--energy insertions would exist in
these circumstances as well. Work along these lines is currently
in progress.
\section{Acknowledgments}
\label{sec6}
E.~P.~thanks with pleasure R.D.~Pisarski for valuable discussions and
the High Energy
Theory group at Brookhaven for their generous hospitality.
R.~K. and M.~C.~thank P.~Aurenche, F.~Gelis, and A.~Smilga for
valuable discussions.
This work was supported by the Natural Sciences and Engineering Research
Council of Canada.
%
%

\clearpage
\end{document}